\documentclass[%
 reprint,
superscriptaddress,
 showkeys,
 amsmath,amssymb,
 aps,
]{revtex4-2}

\usepackage{graphicx}
\usepackage{dcolumn}
\usepackage{bm}
\usepackage{gensymb}
\usepackage{subscript}
\usepackage{tikz}
\usepackage{nicefrac}

\usepackage[textsize=tiny]{todonotes}
\usepackage[normalem]{ulem}

\renewcommand{\eqref}[1]{Eq.~(\ref{#1})}
\newcommand{\eqsref}[1]{Eqs.~(\ref{#1})}
\newcommand{\figref}[1]{Fig.~\ref{#1}}

\newcommand{\appref}[1]{Appendix~\ref{#1}}

\begin{document}

\title{The underappreciated role of nonspecific interactions in the crystallization of DNA-coated colloids}
\author{Hunter Seyforth}
\affiliation{%
 Martin A. Fisher School of Physics, Brandeis University, Waltham, MA 02453 USA\\
}%
\author{Sambarta Chatterjee}
\affiliation{%
 Department of Chemistry, Princeton University, Princeton, NJ 08544 USA\\
}%
\author{Thomas E. Videb\ae k}
\affiliation{%
 Martin A. Fisher School of Physics, Brandeis University, Waltham, MA 02453 USA\\
}%
\author{Manodeep Mondal}
\affiliation{%
 Martin A. Fisher School of Physics, Brandeis University, Waltham, MA 02453 USA\\
}%
\author{William M. Jacobs}%
 \email{wjacobs@princeton.edu}
\affiliation{%
 Department of Chemistry, Princeton University, Princeton, NJ 08544 USA\\
}%
\author{W. Benjamin Rogers}%
 \email{wrogers@brandeis.edu}
\affiliation{%
 Martin A. Fisher School of Physics, Brandeis University, Waltham, MA 02453 USA\\
}%


\begin{abstract}
Over the last decade, the field of programmable self-assembly has seen an explosion in the diversity of crystal lattices that can be synthesized from DNA-coated colloidal nanometer- and micrometer-scale particles. The prevailing wisdom has been that a particular crystal structure can be targeted by designing the DNA-mediated interactions, to enforce binding between specific particle pairs, and the particle diameters, to control the packing of the various species. In this article, we show that other ubiquitous nonspecific interactions can play equally important roles in determining the relative stability of different crystal polymorphs and therefore what crystal structure is most likely to form in an experiment. For a binary mixture of same-sized DNA-coated colloidal micrometer-scale particles, we show how changing the magnitudes of nonspecific steric and van der Waals interactions gives rise to a family of binary body-centered tetragonal crystals, including both cesium-chloride and copper-gold crystals. Simulations using pair potentials that account for these interactions reproduce our experimental observations quantitatively, and a theoretical model reveals how a subtle balance between specific and nonspecific forces determines the equilibrium crystal structure. These results highlight the importance of accounting for nonspecific interactions in the crystal-engineering design process.
\end{abstract} 

\keywords{Crystallization | Programmable self-assembly | DNA-coated colloids | van der Waals}

\maketitle

\section{Introduction}
Self-assembly is a powerful approach for making materials with complex microstructures, whereby individual components spontaneously organize due to interparticle interactions~\cite{whitesides2002self}. In recent years, DNA has emerged as a useful tool to design and control the interactions between colloidal particles owing to the sequence-specific nature of DNA base pairing~\cite{mirkin1996dna,alivisatos1996organization,jacobs2024assembly}. This approach has been used to assemble a rich diversity of crystal lattices from both DNA-coated nanoparticles~\cite{ Mirkin2008Nature, Oleg2008Nature, macfarlane2011nanoparticle, laramy2020crystal} and DNA-coated micrometer-scale particles~\cite{biancaniello2005colloidal,wang2015crystallization,he2020colloidal,hensley2022self,hensley2023macroscopic}. Across this wide spectrum of length scales and crystal structures, the conventional approach has focused on controlling the specific, DNA-mediated attractions~\cite{rogers2011direct,cui2022comprehensive}, along with the particle diameters, in order to specify the formation of a given assembly or stabilize a particular crystal lattice.

Beyond the specific DNA-mediated attractions, however, other surface forces can in principle also play a crucial role in dictating the outcome of colloidal crystallization. Prominent examples include screened-Coulomb repulsion between like-charged particles, steric repulsion between polymer-grafted particles (including DNA-coated particles), and van der Waals (vdW) attraction between particles dispersed in a solvent~\cite{liang2007interaction,israelachvili2011intermolecular}. Although these nonspecific interactions are often overlooked, they can play equally important roles in stabilizing a certain crystal polymorph, especially in cases where there are multiple competing crystal structures with similar free energies. Indeed, recent reports demonstrate how rationally controlling the steric repulsion between DNA-coated nanoparticles can lead to the formation of different crystals~\cite{mao2023regulating}, including crystals that have been notoriously difficult to assemble, like those isostructural to sodium chloride (NaCl) ~\cite{macfarlane2011nanoparticle,wang2015crystallization}.

In this paper, we show how nonspecific interactions affect the assembly of micrometer-scale DNA-coated colloids. In particular, we show that the interplay between specific DNA-mediated attraction, nonspecific vdW attraction, and nonspecific steric repulsion favors the self-assembly of body-centered tetragonal (BCT) crystals, rather than crystals with cubic unit cells. We find that the BCT unit-cell dimensions depend sensitively on the relative magnitude and range of the nonspecific interactions, which we tune by varying the molecular weight of the polymer brush on the particle surface, the particle diameter, and the mixing of complementary grafted DNA strands. Using simulations, we confirm that pair potentials that account for these design choices give rise to equilibrium BCT crystals in quantitative agreement with experiments. Then, to understand these observations, we develop a theory to predict the relative stability of different crystal lattices for particles with both specific and nonspecific attraction. Taken together, our results showcase how nonspecific interactions can play an important role in dictating the equilibrium crystal structures, as well as how they can be controlled by tuning various experimental parameters. Our results might also help to explain why seemingly similar experimental systems produce different crystal structures~\cite{wang2015crystallization,wang2015synthetic,hensley2022self,hensley2023macroscopic}, and why some predicted structures have remained inaccessible~\cite{tkachenko2016generic}. 

\section{Results and Discussion}

\subsection{Characterizing the crystal structures for various polymer-brush lengths}
\begin{figure*}
    \centering
    \includegraphics[width=\linewidth]{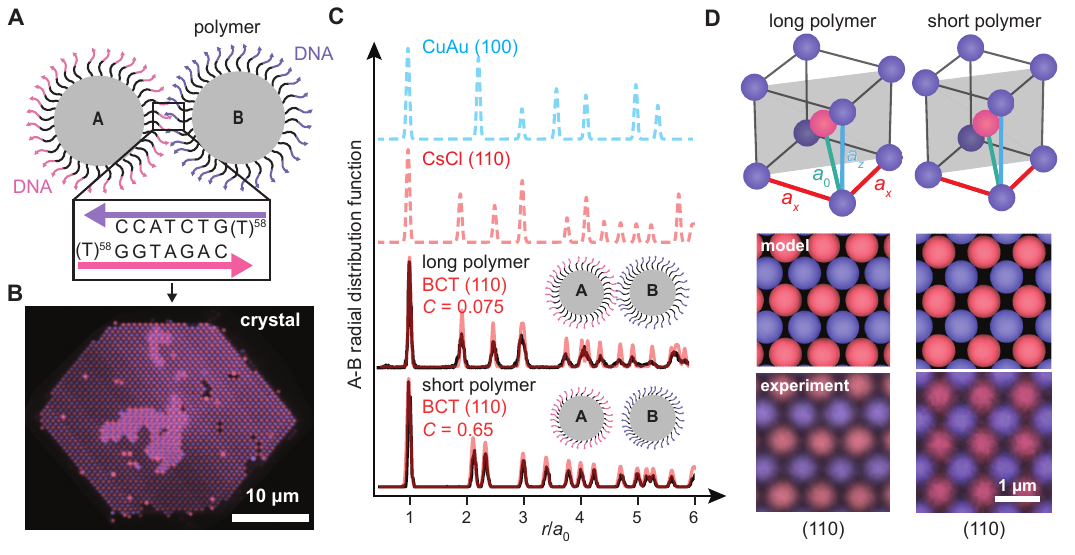}
    \caption{ \textbf{Overview of the experimental system.} (\textbf{A})~Our experimental system consists of 600-nm-diameter spherical colloidal particles that are coated with a brush of polyethylene oxide to which 65-nucleotide-long single-stranded DNA is attached (not to scale).  The DNA terminates in a seven-base-long region designed to hybridize with its complement. Two species of particles (A and B) are defined by the sequences of DNA attached to their surfaces. (\textbf{B})~An equimolar mixture of A and B particles self-assembles to form binary crystals, which we characterize using confocal fluorescence microscopy. (\textbf{C})~The A--B radial distribution functions (RDF) of facets of single crystals assembled with a 6.5$\,$kDa molecular weight ($M_{\text{W}}$) polymer brush and a 67$\,$kDa $M_{\text{W}}$ polymer brush are compared to the RDFs of the closest-matching planes of cesium-chloride (CsCl) and copper-gold (CuAu) crystals, indicated by their miller indices. The distance between particle centers is denoted by $r$, and $a_0$ is the distance between nearest-neighbor A-B particle pairs. Black curves show experimental data. The particles with shorter polymers assemble into a BCT structure in between CsCl and CuAu, whereas the particles with longer polymers assemble into a structure that closely resembles CsCl. Solid red curves show the RDFs for the closest matching BCT crystallographic planes (110), as well as the corresponding $C$ values. (\textbf{D})~By analyzing the A--B radial distribution functions, we generate the unit cells of the lattice structures assembled in experiment. Images of the surfaces of the crystals with long and short polymers are compared to the best-fit model lattices determined from the RDFs. Both examples correspond to (110) planes of BCT.} 
    \label{fig:intro}
\end{figure*}

We synthesize DNA-coated colloids with a range of polymer-brush molecular weights using a combination of physical grafting and click chemistry by following a modified version of Ref.~\cite{oh2015high} (see Supporting Information for more details). Initially, we use polystyrene colloidal particles with 600$\,$nm diameters and four different polyethylene oxide (PEO) block molecular weights, $M_{\text{W}} =$ 6.5, 11, 34, and 67 kDa. We fluorescently dye the two particle species (referred to as A and B) different colors to distinguish their compositional order within the crystal phase. The ssDNA sequence is 65 bases long with a 7 base-long region at the end that is complementary between the two sequences. The net result is a binary suspension of same-sized particles coated with a dense brush of PEO polymers, which are end-grafted with single-stranded DNA~(\figref{fig:intro}A). 

We characterize the binary crystals that self-assemble by determining the two-dimensional radial distribution functions (RDFs) of their surface crystallographic planes. We first assemble crystals at a constant temperature just below their melting point (see Supporting Information for a detailed protocol).
We then lower the sample to room temperature at a ramp rate of 0.05~\degree C every two hours, so that we can transfer it to a confocal fluorescence microscope to image the crystal symmetry and composition.
Next, we characterize the structures of roughly one hundred single crystals (\figref{fig:intro}B) by finding the particle centers using standard image-analysis routines~\cite{crocker1996methods} and computing the two-dimensional RDFs. Two examples for crystals with the longest and shortest polymer lengths are shown in \figref{fig:intro}C. Because the particles are neither index matched nor density matched to the solvent, we image the surface facet that is parallel to the coverslip--water interface. 

We classify the structure of each crystal by examining the measured RDF for A-B particle pairs, $g_{AB}(r)$, which provides a unique fingerprint. Specifically, we compare the measured RDF to a look-up table of RDFs for low-index crystallographic planes of body-centered tetragonal crystals. The look-up table comprises a variety of binary body-centered tetragonal crystals with various aspect-ratio unit cells, which we characterize by a dimensionless parameter $C$ that lies between 0 and 1. If $C=0$, the crystal has a cubic unit cell that is isostructural to cesium chloride (CsCl); if $C=1$, the crystal again has a cubic unit cell, but this time is isostructural to copper gold (CuAu). Intermediate values of $C$ correspond to crystals  with non-cubic unit cells that are characterized by a nearest-neighbor A--B separation distance $a_{0}$ and primitive vectors $a_x\hat{x}$, $a_y\hat{y}$, and $a_z\hat{z}$: 
\begin{subequations}
  \label{eq:a}
\begin{eqnarray}
{a_0^* = \frac{\sqrt{3} a_0}{\sqrt{3}+(2\sqrt{3}-2-\sqrt{2})(C^2-C)}}\\
{a_x=a_y=\frac{2}{\sqrt{3}}a_0^*(1-C) + a_0^* C }\\
{a_z = \frac{2}{\sqrt{3}}a_0^*(1-C) + \sqrt{2}a_0^* C.
}\end{eqnarray}
\end{subequations} 

Interestingly, the crystals that we assemble are neither a perfect match for CsCl nor CuAu. Moreover, their crystal structure depends on the polymer brush molecular weight $M_{\text{W}}$. For example, \figref{fig:intro}C shows the measured RDFs for binary crystals composed of 600-nm-diameter particles with $M_{\text{W}} =$ 6.5~kDa and 67~kDa, as well as reference RDFs for the closest matching crystallographic planes of CsCl and CuAu. By visual inspection, we can clearly see that the two experimental RDFs are different from one another and from the reference RDFs. The differences between the experimentally determined structures are also evident in the crystal images shown in \figref{fig:intro}D. In this way, we find that all four crystals correspond to different BCT lattices with different unit-cell aspect ratios. For the particles with the shortest polymer brush, we also assemble and characterize the crystal structure for three salt concentrations ranging from roughly 50--500~mM NaCl, which has a minimal effect on the equilibrium $C$ parameter value (see Figs.~S1 and S2 in \textit{Supplementary Information}).

We hypothesize that the BCT crystal structures that we observe are stabilized by a combination of specific DNA-mediated attraction and nonspecific interactions. 
Prior theoretical and simulation studies using simple models of DNA-coated particles, which do not account for potential nonspecific interactions present in our experimental system, have predicted that CsCl is the thermodynamically stable crystal phase for same-sized particles with attractive interactions between A and B particles only~\cite{tkachenko2016generic,casey2012driving}. 
This prediction is due to the greater entropy of hard-sphere CsCl crystals, which have body-centered cubic (BCC) unit cells, relative to hard-sphere CuAu crystals, which have face-centered cubic (FCC) unit cells~\cite{hu2018entropy}.
However, this predicted entropic difference between CsCl and CuAu structures is smaller than the thermal energy per particle.
It is therefore plausible that other interactions with similar or greater magnitude can influence the thermodynamically stable crystal structure. 
For example, earlier experimental results showed that CuAu crystals can be stabilized relative to CsCl by introducing a small degree of DNA hybridization between particles of the same type~\cite{casey2012driving}, which can only form contacts in the denser CuAu phase due to the narrow range of distances over which the DNA can hybridize.
Our hypothesis follows similar reasoning, but differs in that the range of the various nonspecific interactions need not be the same as that of the specific DNA-mediated attractions.
We propose that the sum total of these various surface forces, acting over a variety of different ranges, could plausibly stabilize non-cubic BCT crystal structures, which have different nearest-neighbor distances between A-A/B-B and A-B particle pairs.

\begin{figure}
    \centering
    \includegraphics[width=\linewidth]{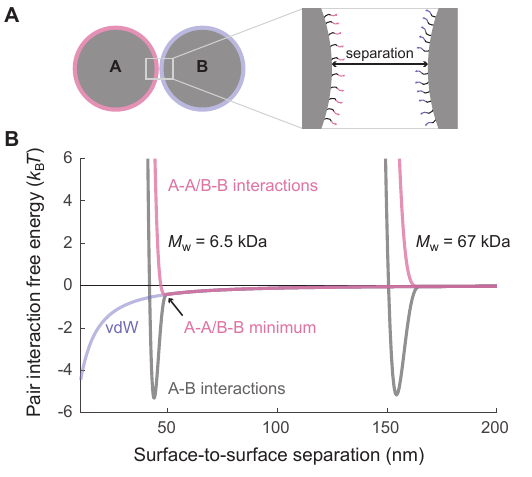}
    \caption{ \textbf{Predicting A-A/B-B and A-B pair potentials.} (\textbf{A})~A schematic of a pair of DNA-coated particles, illustrating the surface-to-surface separation distance. (\textbf{B})~The total A-A/B-B and A-B pair potentials are calculated for the systems examined in Fig.~1C,D (600$\,$nm-diameter particles with $M_{\text{W}} =$ 6.5 or 67$\,$kDa) plotted with respect to the separation distance between the colloidal particle surfaces (black curves).  Because the DNA-hybridization contribution to the pair potential is extremely temperature sensitive, we illustrate the differences between these systems by choosing the temperature separately for each system such that the minimum of the A--B potential is approximately $-5 k_{\text{B}}T$. The blue curve shows the nonspecific vdW contribution to the total pair potentials.  In the $M_{\text{W}} =$ 6.5$\,$kDa system, the vdW attraction is approximately $0.5~k_{\text{B}}T$ at the distance where the total A--B pair potential is minimized.  In the $M_{\text{W}} =$ 67$\,$kDa system, the longer brush screens this interaction.} 
    \label{fig:A-B A-A pair potential}
\end{figure}

\subsection{Modeling microscopic interactions between DNA-coated colloids}
To test this hypothesis, we first evaluate the pair-interaction free energies (referred to as `pair potentials') in our experimental system using a recently developed microscopic model of the interactions between DNA-coated colloids~\cite{cui2022comprehensive}. This microscopic model accounts for the free energy of DNA hybridization~\cite{santalucia2004thermodynamics}, steric repulsion between polymer brushes, and vdW attraction between colloidal particles. Importantly, the model self-consistently accounts for the excluded-volume interactions between adjacent strands, which may or may not be hybridized, by minimizing the total free energy with respect to the fraction of hybridized strands. In this way, the total free energy of interaction can be computed as a function of the distance between the DNA-coated particles. For example, \figref{fig:A-B A-A pair potential} shows predictions of the A-A/B-B and A-B pair potentials for the two systems investigated in \figref{fig:intro}C,D at temperatures near the melting points of the corresponding crystals. Notably, each A-B pair potential has a narrow attractive well when the particle surfaces are separated by a distance comparable to twice the equilibrium polymer brush height. By contrast, the contribution to the pair potentials due to vdW attraction acts over longer distances.

These predicted pair potentials suggest that our hypothesis is indeed plausible. First, we observe that the contribution to the pair potential from vdW attraction is predicted to be independent of $M_{\text{W}}$, as the vdW interaction is determined by the colloidal particle diameter and the Hamaker constant in Ref.~\cite{cui2022comprehensive}. Therefore, the vdW interaction is the same for both A-A/B-B and A-B particle interactions. By contrast, the steric contribution from the polymer brush and the grafted DNA molecules depends sensitively on the polymer $M_{\text{W}}$, whereby longer polymers shift the minimum of the pair potential to larger particle separations. These differences mean that the nonspecific vdW attraction is comparable to the thermal energy for both A-A/B-B and A-B particle pairs when the brushes come into contact in the short-polymer (6.5~kDa $M_{\text{W}}$) system, whereas the vdW attraction is negligible when the brushes come into contact in the long-polymer (67~kDa $M_{\text{W}}$) system. Indeed, this weak A-A/B-B attraction between DNA-coated colloids that depends on the polymer molecular weight has been measured directly for a very similar experimental system~\cite{cui2022comprehensive}.

Second, we observe that the steric repulsion due to excluded volume interactions between the brushes occurs at different distances for the A-A/B-B and A-B particle pairs.
This feature arises because the minimum-free-energy brush height in the microscopic model of Ref.~\cite{cui2022comprehensive} depends on whether the grafted DNA is hybridized.
Because the A-A/B-B particle pairs do not experience DNA-mediated attraction, the repulsion between their polymer brushes leads the pair potential to diverge at a greater separation between the particles.
As we discuss below, this feature also has significant implications for the equilibrium binary crystal structure.

\subsection{Comparing predicted and experimental BCT crystal structures}
We next assess whether these pair potentials can explain the variations in the BCT structures that are observed in experiment.
To this end, we use constant-pressure molecular dynamics simulations to predict the equilibrium crystal structure for each experimental system at room temperature, $T = 25^\circ$C (see \appref{app:sim}).
For each polymer length, we prepare a 1024-particle binary crystal with initial BCT $C$ parameters ranging from 0 to 1.
We then set the applied pressure to be close to zero, since the coexisting colloidal gas phase in the experiments is extremely dilute, and evolve the system under the pair potential that is predicted for the experimental design parameters.
Regardless of the initial $C$ parameter, we find that the structure converges to a unique crystal lattice for each system, which we characterize by computing the time-averaged unit-cell dimensions $a_x$, $a_y$, and $a_z$.

\begin{figure}
    \centering
    \includegraphics[width=\linewidth]{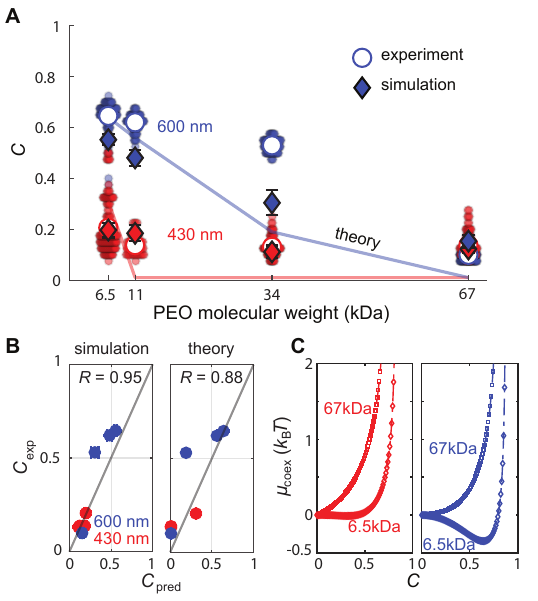}
    \caption{\textbf{Comparison of predicted and experimentally determined BCT crystal structures.}
      (\textbf{A})~A plot of the $C$ parameter values for 430-nm-diameter (red) and 600-nm-diameter (blue) colloids versus the PEO \textit{M}\textsubscript{w} in experiment (circles), simulation (diamonds), and theory (curves). Small circles show data for individual crystals and large white circles show the mean value.
      (\textbf{B})~Plots of the measured $C$ values at room temperature \textit{C}\textsubscript{exp} versus the predicted equilibrium $C$ values from simulation (left) and theory (right) show the strong correlation between experimentally assembled crystals and predicted equilibrium crystal structures. Pearson coefficients $R$ are shown for both comparisons with simulations (left) and theory (right).
      (\textbf{C})~A plot of the chemical potential $\mu$ of BCT crystals for 600-nm and 430-nm-diameter particles as a function of $C$ for two PEO molecular weights: 6.5~kDa (rhombic dashed curve) and 67~kDa (square dashed curve). The chemical potential for the 6.5-kDa case has a minimum at roughly $C=0.7$, illustrating that the equilibrium crystal structure has a non-cubic BCT unit cell, in line with experimental observations.}
    \label{fig:Comparison of Model to Experiment}
\end{figure}

Consistent with our experimental observations, these simulations predict that BCT crystals with $C$ parameters between 0 and 1 are thermodynamically favored over both CsCl and CuAu.
Moreover, the simulations predict equilibrium $C$ parameters, computed via \eqref{eq:a}, that depend on $M_{\text{W}}$ and agree nearly quantitatively with the experimentally characterized crystals (\figref{fig:Comparison of Model to Experiment}A,B).
Both simulations and experiments show that the $C$ value decreases from roughly 0.6 (more CuAu-like) to 0.2 (more CsCl-like) as $M_{\text{W}}$ increases from 6.5 to 67$\,$kDa.
We observe a relatively narrow distribution of $C$ for each system, indicating that all crystals for a particular system assemble the same BCT structure.
Thus, based on the correspondence with our equilibrium simulations and the reproducibility of our experimental results, we conclude that the experimental $C$ parameters reflect the equilibrium crystal structure at room temperature.

Because the vdW contribution to the pair potential is determined by the particle diameter in the microscopic model, we hypothesize that the nonspecific attraction responsible for high $C$ parameter values can also be eliminated by reducing the size of the colloidal particles.
We therefore perform the same set of experiments and simulations for the same polymer lengths but a smaller particle diameter of 430$\,$nm.
For these systems, the vdW attraction is predicted to be negligible at the distance at which the polymer brushes come into contact, regardless of the polymer $M_{\text{W}}$.
As a result, we expect that the BCT crystal structures will all be close to CsCl.

Simulations and experiments confirm this expectation (\figref{fig:Comparison of Model to Experiment}A,B), as the equilibrium $C$ parameter for the smaller-particle crystals is small ($\lesssim 0.2$) and nearly independent of the polymer $M_{\text{W}}$.
Again we observe relatively narrow distributions of $C$ parameters, except in the case of the shortest polymers, for which the $C$ parameter ranges from 0.1 to 0.5 across different crystals.
Yet overall, the semi-quantitative agreement between simulations and experiments is consistent regardless of the particle diameter.
We attribute small discrepancies between the simulations and the experiments to our assumption that the polymer density is constant across all systems, which may not be the case in reality given that each system is synthesized separately.
Nonetheless, the agreement between simulation and experiment gives us confidence that the experimental crystal structures are governed by equilibrium thermodynamics and a balance of specific and nonspecific interactions that make up the predicted pair potentials.

\subsection{Predicting the stability of BCT crystal polymorphs}
To gain deeper insight into the thermodynamic determinants of the observed crystal structures, we devise a theoretical model with which we can predict the equilibrium phase behavior from the microscopically modeled pair potentials.
Our approach utilizes thermodynamic perturbation theory and a reference binary crystal in which A-B particle pairs are bonded via a harmonic potential.
The harmonic potential $u_{\text{ref}}(r)=\epsilon + (k_{\text{AB}}/2) (r-\sigma)^2$ is obtained from the A-B pair potential in the absence of vdW interactions, $u_{\text{AB}}$, where $r$ is the distance between the centers of the particles, $\sigma$ is the distance at which $u_{\text{AB}}$ is minimized, $\epsilon$ is the value of $u_{\text{AB}}$ at $r = \sigma$, and $k_{\text{AB}}$ is the second derivative of $u_{\text{AB}}$ evaluated at $r=\sigma$.
The free energy of a crystal with a specified BCT $C$ parameter can be calculated exactly for this reference model.
However, nonspecific interactions, which we treat perturbatively, turn out to be crucial for determining the relative free energies of the crystal structures.
Complete details of these calculations are provided in \appref{app:theory}.
In what follows, we describe our approach for calculating the phase behavior and highlight the essential insights that emerge.

Given the predicted pair potentials, we seek the conditions for phase coexistence between a colloidal gas phase and a BCT crystal structure with a variable $C$ parameter.
To this end, we compute the free energies of these phases according to the thermodynamic perturbation theory and then equate their chemical potentials, $\mu$, and pressures, $P$.
At phase coexistence,
\begin{equation}
  \label{eq:coex}
    \mu_{\text{fluid}} (\eta_{\text{fluid}}) = \mu_{\text{crystal}} (P_{\text{fluid}} (\eta_{\text{fluid}});C),
\end{equation}
where $\eta_{\text{fluid}}$ is the packing fraction of the fluid phase.
For simplicity, we assume that the packing fraction of the crystal phase is uniquely determined by the BCT $C$ parameter and is thus independent of the pressure.
We solve \eqref{eq:coex} as a function of the packing fraction for values of $C$ in the range 0 to 1, and we predict the equilibrium crystal phase to be the one with the lowest coexistence chemical potential.
Example plots of the coexistence chemical potential, $\mu_{\text{coex}}$, as a function of $C$ are shown in \figref{fig:Comparison of Model to Experiment}C for the two brush lengths and both particle types considered in \figref{fig:intro}C,D.
In all cases, we find that $\mu_{\text{coex}}$ is a function of $C$ with a single local minimum, which is consistent with the fact that systems initialized with various $C$ values all converge to the same crystal structure in our simulations.

We validate the predictions of our theory by comparing both the crystal melting temperatures and the equilibrium $C$ values at room temperature with experiments.
First, we predict the melting temperature for each system theoretically by finding the temperature-dependent pair potentials at which the dilute-phase packing fraction, $\eta_{\text{fluid}}$, that is in coexistence with the equilibrium BCT crystal is equal to the overall colloidal volume fraction in our experiments, $\sim 0.5\%$.
In this way, we find that the predicted and experimentally determined melting temperatures agree semi-quantitatively, with a Pearson correlation coefficient of $R = 0.98$ and absolute errors less than 2\degree C (see Fig.~S3 in \textit{Supplementary Information}).
The melting temperatures decrease monotonically with respect to increasing polymer $M_{\text{W}}$, resulting in a difference of approximately 10\degree C between $M_{\text{W}} = $ 6.5 and 67 kDa.

Second, we predict the BCT $C$ parameters at room temperature and find qualitative agreement with the experimentally determined values.
Although these predictions are not as accurate as the simulation results that utilize the same pair potentials, most likely due to the approximations involved in the perturbation theory, we obtain a positive correlation of $R = 0.88$ with respect to the experimental values (\figref{fig:Comparison of Model to Experiment}B).
We note that the discrepancy between the predictions and the experimental measurements for the 600-nm particles with a 34-kDa polymer brush is possibly due to the very flat free-energy landscape for this system (see Fig.~S4 in \textit{Supplementary Information}), which makes the predicted $C$ values highly sensitive to potential inaccuracies in the pair-potential model, the approximations used in the BCT crystal theory, or our assumption of a constant polymer density across all systems.
Yet overall, our theory reproduces the observed trends with respect to the particle diameter and the polymer $M_{\text{W}}$.
We can therefore examine the theory to understand the thermodynamic driving forces that select the equilibrium BCT crystal structures.

Specifically, our theory predicts that the coexistence chemical potential is affected by the harmonic and anharmonic contributions to the crystal entropy, as well as the vdW attraction.
The harmonic contribution to the entropy, which arises from the phonon modes of the reference crystal, actually favors high $C$ parameter values, although the difference between the CuAu and CsCl phonon entropies is less than $0.1k_{\text{B}}$ per particle (see Fig.~S5 in \textit{Supplementary Information}).
This observation may be surprising given that BCC crystals are typically expected to be entropically favored over FCC crystals~\cite{hu2018entropy}.
However, we emphasize that the harmonic reference crystal does not consider the repulsion between particles of the same type, which is accounted for by the pair potential $u_{\text{AA}} = u_{\text{BB}}$.
This repulsion instead leads to an anharmonic contribution to the crystal entropy, which indeed favors low-$C$ crystal structures in which particles of the same type are less likely to come into contact.
Importantly, the probability that particles of the same type contact one another for a given $C$ parameter depends on the distance at which $u_{\text{AA}}$ becomes strongly repulsive.
As previously noted, this repulsion occurs at greater distances than the repulsion between particles of different types (\figref{fig:A-B A-A pair potential}), leading to an increased anharmonic entropic penalty that favors low $C$ values.
Finally, the vdW interaction lowers the chemical potential when particles of the same type are closer together, favoring larger $C$ values as expected. 
Although these driving forces are individually weak compared to the thermal energy, shifting the balance among them can have large effects on the crystal structure, or potentially lead to broad distributions of $C$ parameters, when $\mu_{\text{coex}}$ is a weakly varying function of $C$ (\figref{fig:Comparison of Model to Experiment}C).

Following from these considerations, an interesting prediction of our theory is that vdW interactions are in fact not essential to obtain BCT structures, since the harmonic contribution to the entropy also favors high $C$ values.
This prediction is supported by simulations conducted in the absence of vdW attraction (see Fig.~S6 in \textit{Supplementary Information}), highlighting the complexity of crystal-structure prediction when many thermodynamic driving forces of similar magnitudes are involved.
However, testing this prediction experimentally would be challenging due to the interdependence of the anharmonic contribution to the entropy and the vdW interactions, as both of these driving forces are controlled in our system by a limited set of accessible design parameters.

Overall, our theory predicts that the equilibrium BCT crystal structure is primarily determined by the balance between the anharmonic contribution to the crystal entropy and the vdW attraction.
Both of these effects can be considered to be nonspecific since neither is governed by DNA-mediated interactions.
Taken together, these theoretical observations explain how subtle changes to the experimentally accessible system design parameters can simultaneously affect multiple weak driving forces and tune the equilibrium BCT crystal structure.

\subsection{Incorporating DNA-mediated A-A/B-B attraction}

\begin{figure}
    \centering
    \includegraphics[width=\linewidth]{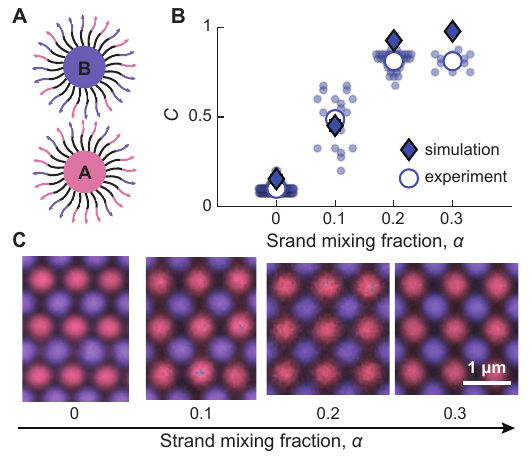}
    \caption{ \textbf{Tuning the BCT crystal structure by balancing specific and nonspecific interactions.}
      (\textbf{A})~600-nm-diameter, $M_{\text{W}} =$ 67$\,$kDa particle types A and B are prepared with different strand mixing fractions, $\alpha$, of two complementary DNA sequences.  (\textbf{B})~Plotting the BCT $C$ parameter versus $\alpha$ reveals a continuous shift towards CuAu-like structures as $\alpha$ increases.  Constant-pressure simulations of the equilibrium crystal structures, conducted with pair potentials that account for the different mixing fractions (diamonds), are in good agreement with the experimental measurements (circles). Small circles show data for individual crystals; large white circles show the mean values.  The broad distribution of experimentally determined $C$ parameters at $\alpha = 0.1$ likely reflects a flat $\mu_{\text{coex}}$ versus $C$ curve (cf.~\figref{fig:Comparison of Model to Experiment}) near the midpoint of the continuous transition, resulting in a wide range of crystal structures with extremely similar stabilities.
      (\textbf{C})~Fluorescence micrographs of the surfaces of the crystals illustrate the continuous shift from a CsCl-like structure at $\alpha = 0$ to a CuAu-like structure for $\alpha \gtrsim 0.2$.}
    \label{fig:Mixing Fraction}
\end{figure}

Based on our understanding of these thermodynamic driving forces, we hypothesize that alternative nonspecific attractive interactions could be used to systematically tune the BCT crystal structure.
To test this idea, we devise an experiment in which the attraction between A-A/B-B particle pairs is due to DNA hybridization rather than vdW attraction. 
Specifically, we use a polymer brush with the largest PEO molecular weight (67~kDa) to screen the vdW attraction. 
We then coat the two particle species with different mixtures of the two complementary DNA sequences to induce DNA-mediated attractions between A-A/B-B particle pairs, such that one particle species is coated with mole fractions $\alpha$ of one sequence and $(1-\alpha)$ of its complement, and the other particle species is coated with $\alpha$ of the complement and $(1-\alpha)$ of the sequence (\figref{fig:Mixing Fraction}A). Therefore, both particle species should experience roughly equal A-A/B-B attraction, and the range of the A-A/B-B attraction should be comparable to the range of the A-B attraction, because both result from hybridization of the same surface-grafted DNA molecules~\cite{fang2020two}.
To perform simulations, we calculate pair potentials by modifying the fraction of complimentary strands on each particle species, such that the average fraction of hybridizing DNA is $2\alpha(1-\alpha)$ on A-A/B-B particle pairs and $\alpha^2 + (1-\alpha)^2$ on A-B particle pairs.
We also utilize these pair potentials in a modified version of our theory to predict the equilibrium BCT $C$ parameter as a function of $\alpha$ (see Fig.~S7 in \textit{Supplementary Information}).

As anticipated, we find that increasing the strand mixing fraction $\alpha$, and therefore the magnitude of the A-A/B-B attraction, induces a transition from CsCl-like to CuAu-like crystals (\figref{fig:Mixing Fraction}B).
The experimentally observed $C$ parameter value increases from approximately 0.1 to 0.8 upon increasing $\alpha$ from 0 to 0.3, and these observations are matched quantitatively by our simulation results over the same range.
Importantly, this transition appears to be continuous, which is consistent with the convex $\mu_{\text{coex}}$ versus $C$ curves that are predicted by our theory (see Fig.~S7 in \textit{Supplementary Information}).
Interestingly, for $\alpha=0.1$, we observe that the interparticle spacing and crystal symmetry appear to vary locally within a single crystal domain (see Fig.~S8 in \textit{Supplementary Information} for an example).
We interpret this heterogeneity as the result of polymorphs with very similar free energies, leading to a large spread in the measured $C$ parameter values at this specific mixing fraction near the midpoint of the transition from low to high average $C$ value (\figref{fig:Mixing Fraction}B).
This observation is also consistent with the continuous transition predicted theoretically, which results in a relatively flat $\mu_{\text{coex}}$ versus $C$ curve, and thus a broad distribution of near-equilibrium crystal structures, near the midpoint of this transition (\figref{fig:Mixing Fraction}B,C). We point out that an earlier study that used mixing of DNA sequences to induce A-A/B-B attraction lacked the spatial resolution and quantitative crystallography to identify this continuous transition~\cite{casey2012driving}.
Nonetheless, we emphasize that the experimentally determined crystals exhibit the same compositional order as the binary crystals without strand mixing for the range of $\alpha$ values that we consider (\figref{fig:Mixing Fraction}B,C).

\section{Conclusions}

Taken together, our results demonstrate how nonspecific colloidal interactions can play  crucial roles in determining the relative stabilities of competing crystal structures. We demonstrate that a binary mixture of same-sized, micron-scale particles can produce a diversity of BCT crystal structures that can be tuned via a variety of physically distinct modes of nonspecific attraction.
In all cases, we theoretically predict a continuous transition, regardless of whether the nonspecific attraction is due primarily to vdW or DNA-mediated forces, which is consistent with the results of simulations and experiments.
These results further support our conclusion that the experimentally determined BCT crystal structures reflect the equilibrium crystal structures at room temperature, and also demonstrate that subtle changes in the thermodynamic driving forces can be accurately predicted by the microscopic model.

We highlight that a thorough examination of the determinants of a continuous transition from BCC to BCT to FCC versus a discontinuous transition directly from BCC to FCC could be an interesting question for future work. In particular, while it is tempting to draw direct comparisons between the phase behavior that we report and the myriad examples of DNA-mediated colloidal crystallization at the nanoscale and microscale~\cite{laramy2020crystal,jacobs2024assembly}, some of which report discontinuous transitions from BCC to FCC~\cite{zhang2015selective,thaner2015entropy}, making these connections is challenging owing to the wide diversity of length scales, particle shapes, conjugation strategies, and DNA motifs used across these many examples. As a result, it is difficult to conclude whether or not the resultant differences in the phase behavior are due to the fact that the vdW attraction is negligibly weak for nanometer-scale particles, that the entropic contributions to the crystal free energies are fundamentally different due to the fact that the grafted DNA molecules are comparable in size to the particle cores, or something else entirely. Determining the mechanistic details of the transitions between different crystal polymorphs therefore requires new dedicated studies designed to control for these many confounding variables. 

From a materials-design perspective, our results point to new possibilities in the programmable crystallization of DNA-coated particles. Whereas the vast majority of the crystal-engineering design rules have focused primarily on the specific DNA-mediated interactions~\cite{Rogers2016,laramy2020crystal}, our work motivates the development of design approaches that include and utilize both specific and nonspecific interactions. Indeed, our experimental results show that including nonspecific attraction, even in simple binary mixtures of same-sized particles, can expand the diversity of crystals that can be self-assembled to include a family of crystal structures with non-cubic unit cells. This observation begs the question: What other crystal structures might be accessible to systems of DNA-coated colloidal particles if one can design both the specific, DNA-mediated attractions and the nonspecific attractions and repulsions?

\begin{acknowledgments}
  This work was supported by the National Science Foundation (DMR-2214590) as well as the Brandeis NSF MRSEC, Bioinspired Soft Materials, DMR-2011846.
  Simulation, analysis, and theoretical modeling scripts are provided at \texttt{https://github.com/wmjac/dnacc-binary-crystal-pub}.
\end{acknowledgments}

\appendix

\section{Constant-pressure simulations}
\label{app:sim}
We perform molecular dynamics simulations using pair potentials computed from the microscopic model of DNA-coated colloidal particles~\cite{cui2022comprehensive} at the experimental temperature $T=25^{\circ}$C.
For each system, we initialize binary crystals consisting of $1024$ colloidal particles ($512$ particles of each type) in $11$ different BCT crystal structures, $C \in [0,0.1,\ldots, 1]$.
The initial lattice spacing is chosen such that the nearest-neighbor A-B particle pairs are separated by the distance at which the pair potential is minimized.
We work in dimensionless units where the colloidal diameter is $d=1$, the reduced simulation temperature is $k_{\text{B}}T=1$, and the masses of the particles are set to $m=1$, so that the time unit is $\tau=d \sqrt{m/k_{\text{B}}T}$.
Because the potential well is extremely narrow, we choose a small timestep equal to $\Delta t=5\times10^{-5} \tau$.
We then perform constant-pressure (NPT) simulations, allowing the orthorhombic simulation box dimensions to fluctuate independently, at a low pressure of $P=10^{-4} k_{\text{B}}T/d^3$.
We note that our results are quantitatively unchanged if the pressure is reduced to $P=10^{-5} k_{\text{B}}T/d^3$.
We run the simulation for $5\times 10^7$ timesteps to ensure that equilibrium is reached.
Since the simulation box dimensions are allowed to fluctuate independently, the BCT $C$ parameter is determined directly from the box dimensions,
\begin{equation}
  C=(a_{\text{max}}-a_{\text{min}}) / \left[ (\sqrt{2}-1)a_0^* \right],
\end{equation}
where $a_{\text{min(max)}}$ is the minimum (maximum) unit cell dimension, $a_0^*$ as defined in Eq.~(1) with $a_0$ calculated as the equilibrium distance between nearest-neighbor A-B particle pairs in simulations.
Finally, we report the BCT $C$ parameter for each system by averaging over the final $10^7$ timesteps of the equilibrated simulation trajectories and over all $11$ initial conditions, which we empirically find to all converge to the same equilibrium crystal structure.
Error bars associated with the equilibrium $C$ values, which are calculated as the standard error of the mean of the $11$ initial conditions, are typically smaller than the symbol sizes in the figures.

\section{Theoretical model of BCT crystal phase coexistence}
\label{app:theory}
\textit{Phase behavior of the reference system.}
We define a reference system based on the pair potential for A-B particle pairs in the absence of vdW interactions, $u_{\text{AB}}$.
Under the conditions of interest, $u_{\text{AB}}$ has a minimum at separation $r = \sigma$, where the free energy is $\epsilon \equiv u_{\text{AB}}(\sigma)$.
For small perturbations in the particle positions, $u_{\text{AB}}$ can be approximated by a harmonic potential $u_{\text{ref}}(r) = \epsilon + (\nicefrac{k_{\text{AB}}}{2}) (r - \sigma)^2$, where $k_{\text{AB}}$ is the second derivative of $u_{\text{AB}}$ evaluated at $r = \sigma$.

In an ordered binary crystal phase, we assume that the average spacing between nearest-neighbor particle centers is exactly $r = \sigma$.
The crystal-phase packing fraction is then
\begin{equation}
\eta_{\text{crystal}} = \left(\frac{\pi \sigma^3}{6}\right) \times \left(\frac{2}{a_x a_y a_z}\right),
\end{equation}
where $v \equiv \pi \sigma^3 / 6$ is the volume associated with a particle, and $a_x$, $a_y$, and $a_z$ are the dimensions of the BCT unit cell, which contains two particles.
Using the harmonic approximation $u_{\text{ref}}$, we can compute the free energy per particle in a crystal phase~\cite{elser2014phonon},
\begin{align}
  \label{eq:fharm}
  \beta f_{\text{crystal}}(C) = &\frac{z_{\text{AB}}\beta\epsilon}{2} - \frac{3}{2} \ln \frac{2\pi}{\beta \sigma^2 k_{\text{AB}}} \\
  &+ \frac{1}{2N} \sum_{\{\kappa\}_{N,C}} \log \kappa - k_{\text{B}}^{-1}\Delta S_{\text{symm}}, \nonumber
\end{align}
where $z_{\text{AB}} = 8$ is the coordination number for A-B contacts and $\beta = 1$, since the pair potentials represent the potential of mean force and thus contain the actual temperature dependence of the system.
The first term represents the free energy of the crystal in its equilibrium configuration, where all nearest-neighbor A-B particle pairs are spaced exactly $r = \sigma$ apart.
The second term arises from the entropy change due to confinement by a harmonic pair potential with spring constant $k_{\text{AB}}$.
The third term accounts for the phonon modes of an $N$-particle lattice with BCT parameter $C$, where $\{\kappa\}_{N,C}$ is the set of nonzero eigenvalues of the Hessian matrix of the harmonic reference crystal with unit nearest-neighbor particle spacing and a unit spring constant.
In practice, we numerically compute the Hessian of a lattice with $8\times8\times8$ unit cells and evaluate the eigenvalue spectrum.
Finally, the fourth term, $k_{\text{B}}^{-1}\Delta S_{\text{symm}} = -\log 2$, is a symmetry factor that accounts for the periodic translational order of particles in an ordered binary crystal.
The chemical potential in a crystal phase is simply the Gibbs free energy per particle,
\begin{equation}
  \beta\mu_{\text{crystal}}(P;C) = \beta f_{\text{crystal}}(C) + \frac{\beta P v}{\eta_{\text{crystal}}(C)}.
\end{equation}

In the fluid phase, we use the Carnahan--Starling equation of state for a hard-sphere fluid with a mean-field correction for the short-range attractions between particles of different types,
\begin{align}
  \beta\mu_{\text{fluid}}(\eta) &= \log \eta + \frac{8\eta - 9\eta^2 + 3\eta^3}{(1 - \eta)^3} + 2\Delta_\epsilon \eta \\
  \beta P_{\text{fluid}}(\eta) v &= \frac{\eta + \eta^2 + \eta^3 - \eta^4}{(1 - \eta)^3} + \Delta_\epsilon \eta^2,
\end{align}
where $\Delta_\epsilon \equiv 4\beta\epsilon [(1 + \delta / 2\sigma)^3 - (1 - \delta / 2\sigma)^3]$ and $\delta \equiv \sqrt{\nicefrac{8}{k_{\text{AB}}}}$ is the approximate width of the harmonic well.

At phase coexistence, the chemical potentials and pressures of the fluid and a crystal phase are equal.
Since we are assuming that the packing fraction of a given crystal phase is solely determined by the BCT parameter $C$, we only need to solve \eqref{eq:coex} to find $\mu_{\text{coex}} = \mu_{\text{fluid}} = \mu_{\text{crystal}}$ as a function of $\eta_{\text{fluid}}$.
The equilibrium crystal in coexistence with the fluid phase is the one with the lowest equilibrium chemical potential.

\textit{Accounting for A-A/B-B repulsion and vdW attraction.}
The chemical potential of a BCT crystal structure depends sensitively on both the repulsion between A-A/B-B particle pairs, which are not bonded and thus do not appear in the reference harmonic crystal, and additional nonspecific attractions among the particles.
We first account for the pair potential between particles of the same type, $u_{\text{AA}} = u_{\text{BB}}$, which has a repulsive shoulder near $r \simeq \sigma$.
We assume that this effect makes a negligible contribution in the fluid phase, since the Carnahan--Starling equation of state already accounts for hard-sphere repulsion among all the particles in the fluid.
However, in a crystal phase, the shoulder reduces the accessible volume per particle due to repulsion between A-A/B-B particle pairs.
The change in the entropy per particle is approximately
\begin{align}
  \label{eq:Sanharm}
  k_{\text{B}}^{-1}\Delta S_{\text{AA}} &= \log \frac{V_{\text{acc,AA}}}{V_{\text{acc,ref}}} \\
  &\approx \log \frac{\int_0^{d_{\text{AA}}} dx\, g_{\text{AA,ref}}(x) \exp[-\beta u_{\text{AA}}(x)]}{\int_0^{d_{\text{AA}}} dx\, g_{\text{AA,ref}}(x)}, \nonumber
\end{align}
where $V_{\text{acc,ref}}$ and $V_{\text{acc,AA}}$ are the accessible volumes per particle under the reference and actual A-A/B-B potentials, respectively, and $d_{\text{AA}}$ is the minimum distance between A-A/B-B particle pairs in the crystal structure.
In this approximation, the integrals describe the distance over which each particle can travel before being repelled by a particle of the same type within a lattice, assuming that all other particles are at their equilibrium positions.
For simplicity, we approximate the radial distribution function for A-A/B-B particle pairs in the reference crystal as $g_{\text{AA,ref}}(x) \propto \exp[-(\nicefrac{\alpha k_{\text{AB}}}{2})(x - \nicefrac{d_{\text{AA}}}{2})^2]$, where the scaling factor $\alpha$ is chosen to be $\alpha \approx 1\times 10^{-4}$ based on zero-pressure simulations of the reference crystal with $C=0$.
This choice accounts for the fact that the distance between nonbonded A-A/B-B particle pairs fluctuates to a much greater extent than the distance between bonded A-B particle pairs in the reference crystal.
The entropy change given by \eqref{eq:Sanharm} modifies the chemical potential in the crystal phase, such that $\beta\Delta\mu_{\text{crystal}}^{\text{AA}} = -k_{\text{B}}^{-1}\Delta S_{\text{AA}}$.
We emphasize that this anharmonic contribution to the crystal entropy is essential for stabilizing BCT crystals with low $C$ values: In the absence of A-A/B-B repulsion, the $C$-dependent phonon contribution to \eqref{eq:fharm} actually favors crystal structures with high $C$ values.
Thus, the shoulder of the A-A/B-B pair potential plays an important role in determining the equilibrium crystal structure.

By contrast, nonspecific vdW interactions affect particles in both the fluid and crystal phases.
We treat the pair potential $u_{\text{vdW}}$ as a weak perturbation to the reference crystal structure using the $\lambda$-expansion~\cite{hansen2013theory}.
The corrections to the chemical potential and the pressure of the fluid phase are
\begin{align}
  \label{eq:DmufvdW}
  \beta\Delta\mu_{\text{fluid}}^{\text{vdW}} &= \frac{24\eta}{\sigma^3} \int_{\sigma}^\infty dr\, r^2 \beta u_{\text{vdW}}(r) \\
  \label{eq:DPfvdW}
  \beta \Delta P_{\text{fluid}}^{\text{vdW}} v &= \frac{12\eta}{\sigma^3} \int_{\sigma}^\infty dr\, r^2 \beta u_{\text{vdW}}(r).
\end{align}
For the crystal-phase calculation, we approximate the structure of the reference crystal by assuming that the particles are at their equilibrium positions in the binary lattice.
The correction to the chemical potential of the crystal phase is thus
\begin{equation}
  \label{eq:DmuxvdW}
  \beta\Delta\mu_{\text{crystal}}^{\text{vdW}} = \sum_i \beta u_{\text{vdW}}(r_i),
\end{equation}
where the sum runs over all the particles in the lattice except a tagged particle at the origin, and $r_i$ is the distance between the neighboring particle and the origin.

In practice, we incorporate these correction terms, \eqsref{eq:Sanharm}--(\ref{eq:DmuxvdW}), and solve \eqref{eq:coex} to compute $\mu_{\text{coex}}$ as a function of $C$, as shown in \figref{fig:Comparison of Model to Experiment}C.
An extension of this theory to systems with specific A-A/B-B attraction, which follows a similar perturbative approach, is presented in the \textit{Supplementary Information}.

\end{document}


\onecolumngrid

\section*{Supplementary Information for\\``The underappreciated role of nonspecific interactions in the crystallization of DNA-coated colloids''} 
\section{\label{sec:Materials}Materials and Methods}

\noindent \textbf{Synthesizing DNA-coated colloids}
Colloidal particles are functionalized with single-stranded DNA through click chemistry and physical grafting, following a modified version of the methods described by Pine and co-workers ~\cite{Pine2015ChemMater}. A detailed protocol is provided in the Supporting Information of Ref.~\cite{hensley2023macroscopic}. In brief, we prepare four polystyrene-block-poly(ethylene oxide) (PS-b-PEO) copolymers with roughly constant PS molecular weights and varying PEO molecular weights (6.5~kDa, 11~kDa, 34~kDa, and 67~kDa PEO) that are end-functionalized with an azide group (Polymer Source, Inc; part nos. P44891-SEO, P11118C-SEO, P1807A-SEO, and P1802-SEO)). The modified PS-b-PEO is physically adsorbed onto the surface of sulfate-modified polystyrene colloidal particles by swelling/deswelling, and a DBCO-modified ssDNA oligonucleotide is attached to the PS-b-PEO via click chemistry. Next, the colloidal particles are washed five times in aqueous buffer containing 10 mM Tris and 1 mM EDTA (1xTE) through centrifugation and resuspension. The colloidal particles are then stored in a refrigerator at 1\% volume fraction in 1xTE.
We study crystallization of a binary mixture of same-sized DNA-coated colloids. One particle species is coated with sequence A (5’-(T)51-GAGTTGCGGTAGAC-3’) and dyed with an equal volume of a 10\%-saturated solution of Pyrromethane 546 in toluene; the other particle species is coated with sequence B (5’-(T)51-AATGCCTGTCTACC-3’) and dyed with an equal volume of 50\%-saturated solution of nile red in toluene. Both DNA sequences are obtained from Integrated DNA Technologies (IDT) and are purified by high-performance liquid chromatography (HPLC). All crystallization experiments are performed in 1xTE buffer with 500~mM NaCl. 
We observed that the melting temperature decreased with increasing PEO molecular weight for both 600-nm-diameter and 430-nm-diameter colloids, which matches the observations from Pine and co-workers ~\cite{cui2022comprehensive}. \\

\noindent \textbf{Crystallization Experiment}
The sample is prepared by mixing equal parts A and B particles. The final sample consists of 0.5\% volume fraction DNA-coated colloids in 1xTE containing 500~mM NaCl. The solution is loaded into a sample chamber, which is prepared by plasma cleaning a 24~mm by 60~mm No 1. coverslip and a 22~mm by 22~mm No 1. coverslip. A box-shaped border of vacuum grease is placed on the larger cover slide. Then 4~$\upmu$L of the sample is placed inside the vacuum grease box. A 22~mm by 22~mm coverslip is placed over the vacuum grease border and excess air is pushed out of the box leaving a stable sample chamber. UV glue is added to the edges of the small coverslip and cured for 20 minutes. 

We conduct bulk nucleation experiments by first holding the sample at a constant temperature just below melting and then annealing the crystals with a temperature ramp of 0.05~\degree C per 2 hours until few colloids in the gas phase remain. Then the sample is brought to room temperature where the crystals remain stable. The sample temperature is controlled by a thermoelectric cooler. Details of the experimental setup are provided in Refs.~\cite{hensley2022self,hensley2023macroscopic} \\

\noindent \textbf{Microscopy and particle centroiding} \\
Bright-field microscopy images are obtained using a Nikon Ti2 microscope with a 10x-magnification, 0.45 NA objective (MRD00105, Nikon), a 1.5x-magnification tube lens, and a Pixelink M12 Monochrome camera (M12M-CYL, Pixelink) connected to a desktop computer. 
We image the surface of crystals with a Leica SP8 laser-scanning confocal microscope. 

We measure the particle centroids using standard image analysis methods~\cite{crocker1996methods}. In brief, the images are split into separate channels corresponding to the two particle types in the crystal. Each image is convolved with a Gaussian to reduce the noise. A local thresholding algorithm is used to isolate the center of particles and to compensate for any large-scale variations in image intensity. Lastly, we compute the centroids to find the positions of all the particles in the facet. The positions of the particles are then used to calculate the experimental radial distribution function, $g(r)$~\cite{chandler}.

\section{Solving the crystal structures}

We use a quantitative pipeline based on laser-scanning confocal microscopy, image analysis, and crystallography to determine the 3D crystal structures that we explore in experiment, previously described in Ref.~\cite{hensley2023macroscopic}. The pipeline is as follows: (i) collect images of the crystal facets that have sedimented during growth; (ii) find particle positions for both A and B particles using image analysis and compute their radial distribution functions between A-B particle pairs, $g_{AB}(r)$, and A-A/B-B particle pairs, $g_{AA}(r)$; (iii) generate a look-up table of model $\Tilde{g}(r)$ for potential crystalline facets (from here on $\Tilde{g}$ will denote model functions); and (iv) compare the experimental $g(r)$ against all model $\Tilde{g}(r)$ to find the closest match. 

To create the reference $\Tilde{g}(r)$ data, we identify relevant facets of proposed crystal structures that may show up in experiment. We then generate position data of A and B particles for those facets given a proposed crystal structure. In our experiments, we primarily see the (110) and (101) planes of body-centered tetragonal (BCT) structures. Once the particle positions are defined, we calculate corresponding $g(r)$ data sets. In our lookup table we included the (100), (110), (111), and (210) facets of binary BCC (CsCl) and FCC (CuAu) structures, as well as the (110), (101) facets of BCT structures with $C$-values ranging from 0 to 1 in steps of 0.025. Other facets, such as the (100) or (111) facets of BCT, were excluded because we do not observe facets that contain only a single particle type.

Next, we compare the experimental and model $g(r)$ data to identify the best match.  In order to make this comparison, we need to introduce noise into our model data that is comparable to the noise in our experimental data. We fit the $g(r)$ peak that is closest to $r/\sigma=1$ (2 for $g_\mathrm{AA}(r)$) with a Gaussian and then extract the mean and standard deviation; $\sigma$ denotes the minimum A-B particle spacing. The mean is used to get the true lattice spacing of the crystal and to rescale the experimental $g(r)$ to have a peak at $r/\sigma$ at 1 (2 for $g_\mathrm{AA}(r)$). We then convolve the model data with a Gaussian that has the same standard deviation as the experiment. Both sets of data are then normalized to set the amplitude of the first peak to one. This normalization sets a reference point from which we can compare various model facets to our experiment. For a given model $\Tilde{g}(r)$, we compute the sum of the squared residuals between the model and the experimental data, $\sum_i( g_(r_i)-\Tilde{g}(r_i))^2$. We take the model with the smallest sum of the squared residuals as the best fit crystal structure to the experimental image. 

\section{Further development of the theoretical model}

\subsection{Extension of the theoretical model to systems with specific A-A/B-B attraction}

In systems where both A-A/B-B and A-B particle pairs interact via specific attractions, we assume that the A-B interaction is substantially stronger, so that the ordered binary harmonic crystal remains an appropriate reference system.
We then treat the repulsive and attractive portions of the A-A/B-B pair potential, $u_{\text{AA}} = u_{\text{BB}}$, separately when computing the corrections to the reference crystal free energy.
Specifically, we follow the Weeks--Chandler--Anderson approach~\cite{hansen2013theory} by splitting $u_{\text{AA}}$ into repulsive, $u_{\text{AA}+}$, and attractive, $u_{\text{AA}-}$, parts at the minimum of the AA-particle pair potential $r = \sigma_{\text{AA}}$, where $\epsilon_{\text{AA}} = \min u_{\text{AA}} = u_{\text{AA}}(\sigma_{\text{AA}})$:
\begin{align}
  u_{\text{AA}+}(r) &= u_{\text{AA}}(r) - \epsilon_{\text{AA}} \;\text{if}\; r\le\sigma_{\text{AA}}\;\text{otherwise}\; 0 \\
  u_{\text{AA}-}(r) &= \epsilon_{\text{AA}} \;\text{if}\; r\le\sigma_{\text{AA}}\;\text{otherwise}\; u_{\text{AA}}(r).
\end{align}
The repulsive part, $u_{\text{AA}+}$, substitutes for $u_{\text{AA}}$ in Eq.~(B6) of the main text, whereas the attractive part, $u_{\text{AA}-}$, is used to compute the corrections $\Delta\mu_{\text{fluid}}^{\text{AA}-}$ and $\Delta P_{\text{fluid}}^{\text{AA}-}$ following the $\lambda$-expansion as in App.~B of the main text, while accounting for the fact that this correction only affects A-A/B-B particle pairs,
\begin{align}
  \beta\Delta\mu_{\text{fluid}}^{\text{AA}-} &= \frac{12\eta}{\sigma^3} \int_{\sigma}^\infty dr\, r^2 \beta u_{\text{AA}-}(r) \\
  \beta \Delta P_{\text{fluid}}^{\text{AA}-} v &= \frac{6\eta}{\sigma^3} \int_{\sigma}^\infty dr\, r^2 \beta u_{\text{AA}-}(r).
\end{align}
Since the A-A/B-B attraction is short-ranged, we use the approximate radial distribution function for A-A/B-B particle pairs, $g_{\text{AA,ref}}$, as in Eq.~(B6) of the main text when applying the $\lambda$-expansion to $u_{\text{AA}-}$ to calculate $\Delta\mu_{\text{crystal}}^{\text{AA}-}$,
\begin{equation}
  \beta\Delta\mu_{\text{crystal}}^{\text{AA}-} = z_{\text{AA}} \int_0^\infty dx\, g_{\text{AA,ref}}(x) \beta u_{\text{AA}-}(x),
\end{equation}
where $z_{\text{AA}} = 4$ is the coordination number of A-A contacts and the integration range is extended to infinity to capture the complete range over which $g_{\text{AA,ref}}$ is nonzero.
We note that, in principle, the attractive part, $u_{\text{AA}-}$, must be a weak perturbation relative to $\beta^{-1}$ to apply the $\lambda$-expansion.
Although this is generally not the case for the systems and real temperatures that we consider, we nonetheless expect that this approach will predict the correct qualitative effects of specific A-A/B-B attraction.
Representative results, corresponding to Fig.~4 in the main text, are shown in \figref{SIfig:theory-like}.

\clearpage
\section{Additional figures}

\begin{figure}[h!]
    \centering
    \includegraphics[width=0.5\textwidth]{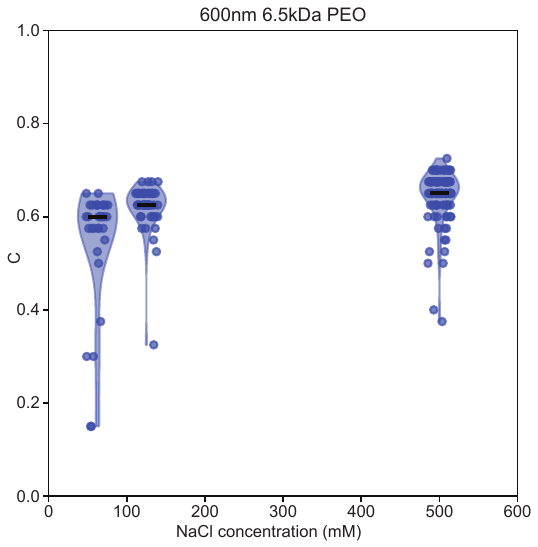}
    \caption{\textbf{Effect of NaCl concentration on the equilibrium $C$ value for 600$\,$nm 6.5$\,$kDa PEO.} Violin plots of $C$ values of crystal structures assembled by 600$\,$nm 6.5$\,$kDa PEO in three different salt conditions show that there is little to no change in the crystal structure through variation of the salt concentration. This observation suggests that there is little to no change in the polymer conformation at the different salt conditions and that electrostatics do not play a dominant role.}

    \label{SIfig:Cval_vary_nacl}
\end{figure} 
 \begin{figure}[h!]
    \centering
    \includegraphics[width=0.5\textwidth]{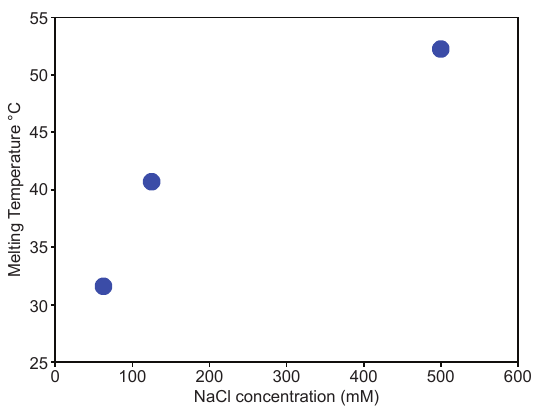}
    \caption{\textbf{Effect of NaCl concentration on the melting temperature.}  The melting temperatures decreases with decreasing salt concentration and approaches room temperature at the lowest salt concentrations for the systems studied in \figref{SIfig:Cval_vary_nacl}.}

    \label{SIfig:T vs NaCl}
\end{figure}

\begin{figure}[h!]
    \centering
    \includegraphics[width=0.5\textwidth]{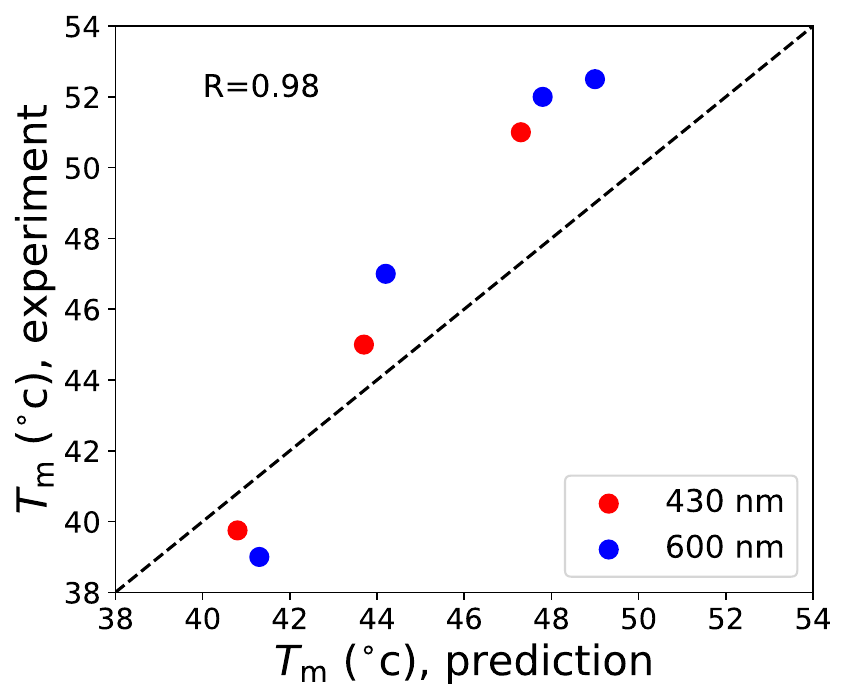}
    \caption{\textbf{Comparison of predicted and experimentally measured melting temperatures.}
      To predict the melting temperature, we find the real temperature $T$ at which the coexistence fluid phase packing fraction, $\eta_{\text{fluid}}(\mu_{\text{coex}})$, is equal to the total packing fraction in the experiments, $0.5\%$.
      This point on the phase diagram corresponds to the melting temperature, $T_{\text{m}}$, since the equilibrium volume fraction occupied by the crystal phase vanishes.
      In general, this occurs when the minimum of the A-B pair potential, $\epsilon$, is approximately $-5 \beta^{-1}$, although the shapes of both the A-B and A-A/B-B pair potentials affect the prediction via the harmonic and anharmonic contributions to the BCT crystal free energy.
      The Pearson correlation coefficient between predictions and experimental measurements is $R = 0.98$.}
    \label{SIfig:melting}
\end{figure}

\begin{figure}[h!]
    \centering
    \includegraphics[width=0.5\textwidth]{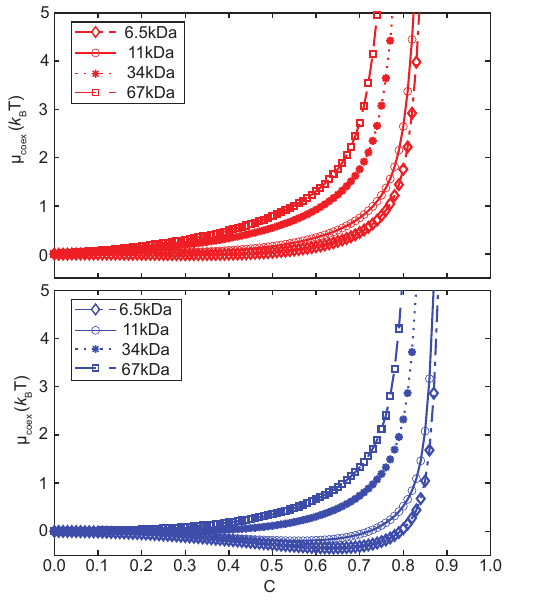}
    \caption{\textbf{Predicted free-energy landscapes for all particle sizes and PEO brushes tested.} Expanded presentation of the $\mu$ versus $C$ curves shown in Fig.~3C in the main text, showing predictions for all the cases explored in Fig.~3A,B.}

    \label{SIfig:Mu_vs_C}
\end{figure} 

\begin{figure}[h!]
    \centering
    \includegraphics[width=0.5\textwidth]{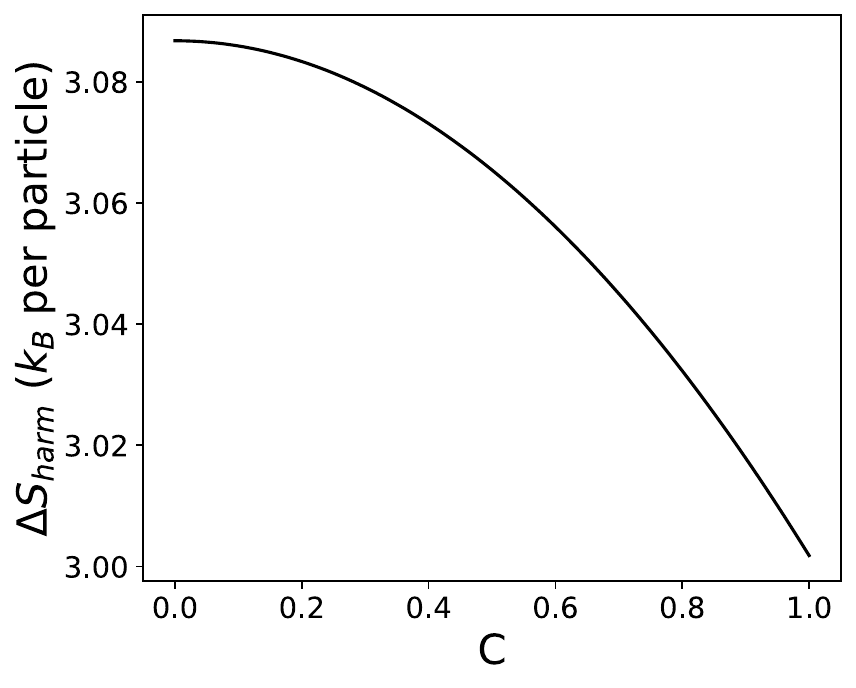}
    \caption{\textbf{Theoretical prediction of the harmonic contribution to the BCT crystal entropy.}  The harmonic contribution to the entropy, $\Delta S_{\text{harm}}$, is given by the second and third terms of Eq.~(B2) in the main text.  Although high $C$ BCT crystals are favored by the harmonic contribution, the difference between CuAu and CsCl crystals of the phonon entropies is less than $0.1k_{\text{B}}$ per particle.}
    \label{SIfig:phonon-entropy}
\end{figure}

\begin{figure}[h!]
    \centering
    \includegraphics[width=0.5\textwidth]{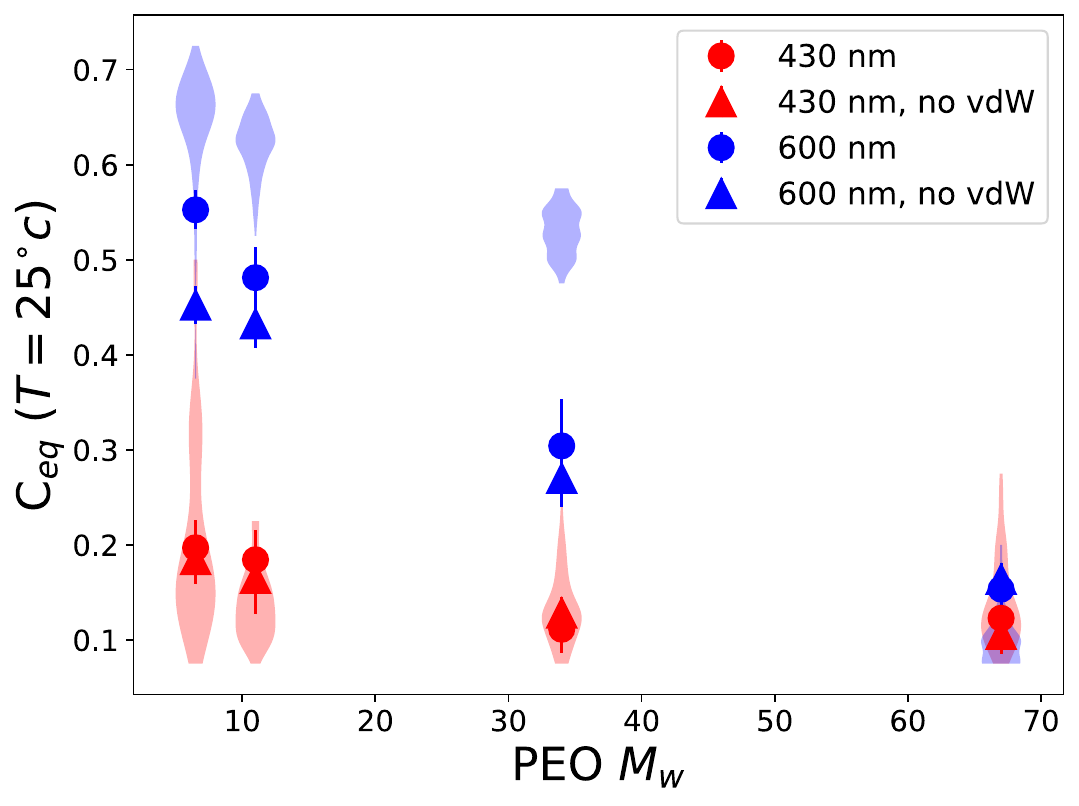}
    \caption{\textbf{Nonspecific vdW attraction is not required to observe high $C$ values.} Simulations predict the equilibrium BCT $C$ parameter in the presence (filled circles) and absence (filled triangles) of nonspecific vdW attraction. For larger particles and shorter polymers, vdW attraction significantly shifts the equilibrium $C$ parameter to larger values.  However, vdW attraction is not required to observe $C$ parameters greater than $\sim 0.2$, as it is not the only contribution to the binary crystal free energy that tends to stabilize BCT crystal structures with larger $C$ values.}
    \label{SIfig:novdW}
\end{figure}

\begin{figure}[h!]
    \centering
    \includegraphics[width=0.9\textwidth]{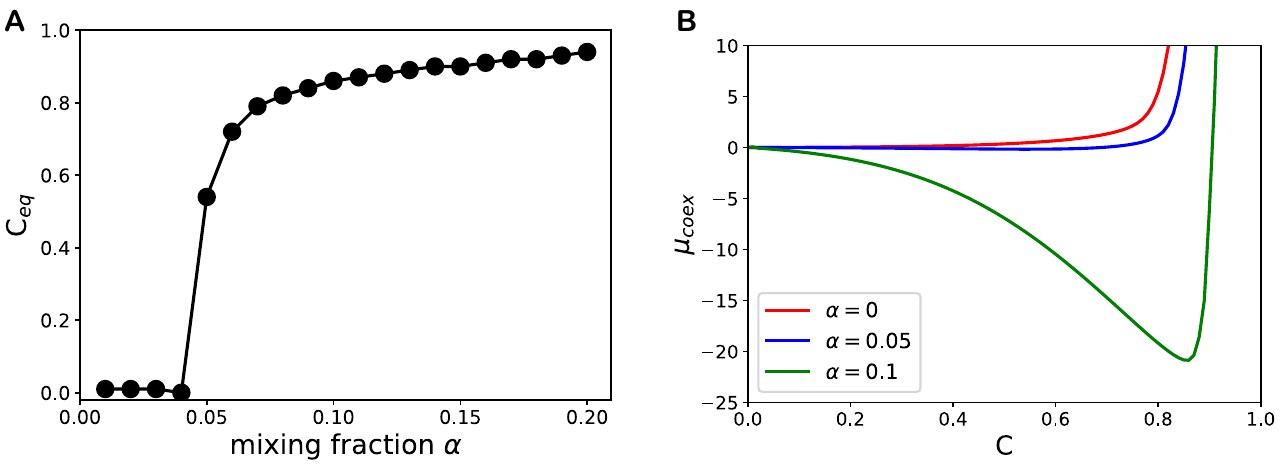}
    \caption{\textbf{Theoretical predictions of the effect of specific A-A/B-B attraction.}
      (\text{A})~The theory predicts a continuous shift of the equilibrium BCT $C$ parameter as the mixing fraction, $\alpha$, is increased.  This trend is consistent with the results of simulations and experimental measurements shown in Fig.~4 of the main text, although the midpoint of the transition is shifted to smaller values of $\alpha$ due to the approximations invoked in the theory.  Calculations are shown here for the 600$\,$nm, $M_{\text{W}} =$ 67$\,$kDa system.
      (\text{B})~Representative $\mu_{\text{coex}}$ versus $C$ curves below, near, and above the midpoint of the theoretically predicted transition.}
    \label{SIfig:theory-like}
\end{figure}

\begin{figure}[h!]
    \centering
    \includegraphics[width=0.5\textwidth]{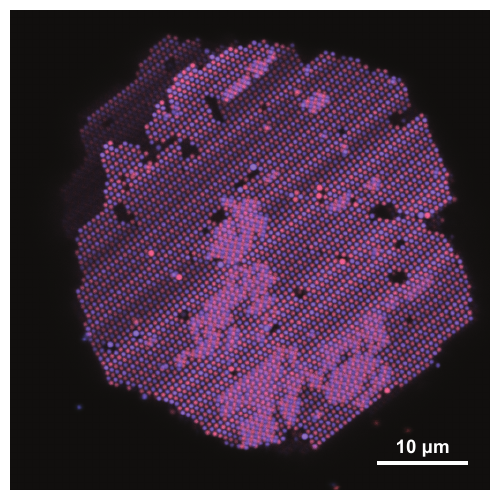}
    \caption{\textbf{Example crystal with spatial variations in the crystal structure.} Confocal fluorescence micrograph of the surface of a crystal assembled from 600$\,$nm particles with a 67$\,$kDa PEO polymer brush and a mixing fraction of $\alpha=0.1$. The crystal packing and symmetry varies from region to region within the crystal. }

    \label{SIfig:alpha=0.1}
\end{figure}